\pdfoutput=1
\documentclass[a4paper,11pt]{article}
\usepackage[margin=2.5cm]{geometry}
\usepackage[utf8]{inputenc}
\usepackage{hyperref}
\usepackage{amsmath,amsfonts,amssymb,array,graphicx,color,caption}
\usepackage{authblk}

\makeatletter

\@addtoreset{equation}{section}
\makeatother

\newcommand{\secref}[1]{Sec.~\ref{sec:#1}}
\newcommand{\figref}[1]{Fig.~\ref{fig:#1}}

\newcommand{\normal}[1]{\ :\! #1 \!\!:}
\newcommand{\aver}[1]{\left\langle {#1} \right\rangle}
\newcommand{\smallaver}[1]{\langle {#1} \rangle}

\newcommand{\vect}[1]{\boldsymbol{#1}}

\newcommand{\nn}{\nonumber}
\newcommand{\zb}{\bar{z}}
\newcommand{\wb}{\bar{w}}
\newcommand{\hb}{\bar{h}}
\newcommand{\om}{\vect\omega}
\newcommand{\vQ}{\vect{Q}}
\newcommand{\ve}{\vect{e}}
\newcommand{\vh}{\vect{h}}
\newcommand{\Zbb}{\mathbb{Z}}
  
\title{Three-point functions in the fully packed loop model on the honeycomb lattice}
\author{T. Dupic}
\author{B. Estienne}
\author{Y. Ikhlef}
\affil{Sorbonne Universit\'e, CNRS, Laboratoire de Physique Th\'eorique et Hautes \'Energies, LPTHE, F-75005 Paris, France}

\begin{document}

\maketitle

\begin{abstract}
  The Fully-Packed Loop (FPL) model on the honeycomb lattice is a critical model of non-intersecting polygons covering the full lattice, and was introduced by Reshetikhin in 1991. Using the two-component Coulomb-Gas approach of Kondev, de Gier and Nienhuis (1996), we argue that the scaling limit consists of two degrees of freedom: a field governed by the imaginary Liouville action, and a free boson. We introduce a family of three-point correlation functions which probe the imaginary Liouville component, and we use transfer-matrix numerical diagonalisation to compute finite-size estimates. We obtain good agreement with our analytical predictions for the universal amplitudes and spatial dependence of these correlation functions. Finally we conjecture that this relation between non-intersecting loop models and the imaginary Liouville theory is in fact quite generic. We give numerical evidence that this relation indeed holds for various loop models. 
  \end{abstract}

\section{Introduction}
\label{sec:intro}

\begin{figure}[ht]
  \centering
    \includegraphics[width=0.3\linewidth]{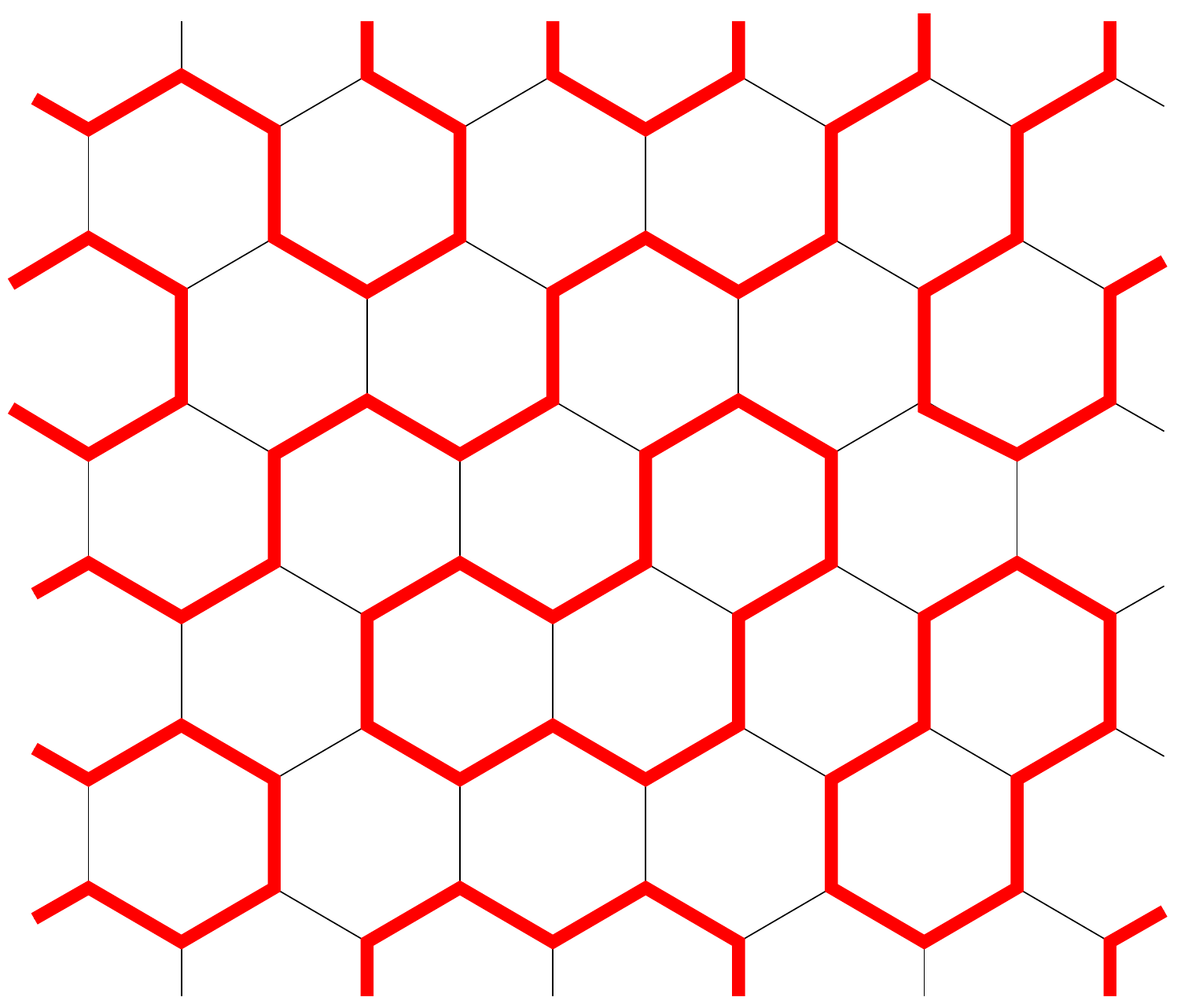}
    \caption{An example configuration of the FPL model on the honeycomb lattice. Loops are drawn in red.
    }
    \label{fig:FPL}
\end{figure}

The Fully-Packed Loop (FPL) model~\cite{reshetikhin91} is a model of non-intersecting closed polygons which cover the hexagonal lattice. It has a single external parameter: the loop fugacity $n$. The partition function is
\begin{equation} \label{eq:Z}
  Z_{\rm FPL} = \sum_{\text{config. $C$}} n^{\# \text{loops  of $C$}} \,,
\end{equation}
where the sum is over all admissible lattice polygon configurations, {\it i.e.} collections of non-intersecting closed polygons covering every site of the lattice: see \figref{FPL}. In particular, the model corresponds to Hamiltonian walks on the hexagonal lattice for $n=0$, dimer coverings for $n=1$, and three-colourings of the hexagonal lattice for $n=2$. 

The Coulomb-Gas (CG) analysis was carried out in~\cite{Kondev-FPL}, and the eigenvalues of the transfer matrix where computed by Bethe Ansatz solution in~\cite{FPL-NLIE}. The model is critical for fugacities $n \in ]-2,2]$, and it is described by a Conformal Field Theory (CFT) of central charge $c$ related to the loop fugacity $n$ by:
\begin{equation} \label{eq:cc}
c = 2 - 6\left( \frac{1}{b} - b \right)^2 \,,   \qquad  n = -2 \cos \pi b^2 \,,
  \qquad \text{with} \quad 0<b \leq 1 \,.
\end{equation}
More recently~\cite{DEI-FPL}, the present authors have revisited the problem of determining the full spectrum of primary operators, which has led to the discovery of excitations with fractional momenta, much like in the standard case of the O($n$) model~\cite{DSZ87}. It was also pointed out that, for a proper modification of the non-contractible loop fugacity on the cylinder, the spectrum can be described in terms of the Kac table of degenerate dimensions for the $W_3$ algebra.

Our general motivation is to make progress towards the description of correlation functions in non-local critical models such as loop models. In particular, the structure constants of the operator algebra are generally related to the amplitudes of three-point functions of primary operators:
\begin{equation} \label{eq:Cabc}
  C_{abc} = \aver{\Phi_a(\infty) \Phi_b(1) \Phi_c(0)} \,.
\end{equation}
In the case of critical loop models, the effective scaling theory typically has an infinite but discrete spectrum of primary operators, which cannot be treated by the standard methods developed for rational CFT~\cite{DotsenkoFateev84,Runkel99}. Interestingly, on the example of critical percolation, it was shown in~\cite{DelfinoViti11} that some three-point amplitudes of CFTs associated to non-local critical models were predicted by the corresponding formula in the Imaginary Liouville (IL) CFT \cite{Schomerus03,Zamo05,KostovPetkova06}, {\it i.e.} the Liouville CFT with an imaginary value of the background charge. Since then, the IL model has received a renewed attention from the point of view of Statistical Mechanics applications \cite{PSVD13,EI15,centralchargeless1,Migliaccio2017}.

In the present paper, we study a family of two- and three-point correlation functions in the FPL model (\ref{eq:Z}--\ref{eq:cc}), as it was done for the O($n$) model in~\cite{IJS16}. These non-local correlation functions are defined by changing the loop fugacity for the topologically non-trivial loops on the Riemann sphere with two or three punctures at some given points. Using a careful CG analysis based on \cite{Kondev-FPL}, where we stress the role of screening charges, we obtain two main results:
\begin{itemize}
\item The conformal dimension associated to a modified loop fugacity $n'=2 \cos 2\pi bp$ with $2bp \in [-1,1]$ is given by $\Delta_p = p^2 - \frac{1}{4}(b^{-1}-b)^2$.
\item The corresponding three-point amplitude $C(n_1,n_2,n_3)$ with $n_j=2\cos 2\pi bp_j$ is given by the three-point function of vertex operators \cite{Schomerus03,Zamo05,KostovPetkova06} in the IL theory with central charge $c_{\rm IL} = c-1$.
\end{itemize}
These results are supported by a numerical analysis based on the exact diagonalisation of the transfer matrix.

More generally this family of two- and three-point correlation functions can be defined for any non-intersecting loop model. Based on numerical evidence, we conjecture that the corresponding amplitudes are universal and do not depend on the particular loop model under consideration. This is quite surprising given that different loop models correspond to different universality classes. 

\newpage 
\section{Coulomb-gas approach}
\label{sec:CG}

\subsection{The Coulomb-Gas action and its parameters}
\label{sec:CG-action}

The CG description of the FPL model, {\it i.e.} the formulation of the scaling limit as a compactified bosonic model coupled to the scalar curvature, and with interaction terms in the form of ``screening charges'', was derived in \cite{Kondev-FPL}. It involves the root lattice $\cal R$ and the weight lattice $\cal R^*$ of the Lie algebra $\mathfrak{sl}_3$:
\begin{equation*}
  \mathcal{R} = \Zbb\ve_1  + \Zbb\ve_2 \,,
  \qquad \mathcal{R}^* = \Zbb \om_1 +  \Zbb\om_2 \,.
\end{equation*}
We use the conventions:
$$\om_1^2=\om_2^2 = \frac{2}{3} \,, \qquad \om_1\cdot\om_2 = \frac{1}{3} \,,$$
and:
$$\ve_1 = 2\om_1-\om_2 \,, \qquad \ve_2=2\om_2-\om_1 \,, \qquad \ve_i \cdot \om_j = \delta_{ij} \,.$$
It will also be convenient to introduce the Weyl vector $\vect\rho=\om_1+\om_2=\ve_1+\ve_2$ with square norm $\vect\rho^2=2$, and the weights of the first fundamental representation:
$$\vh_1 = \om_1 \,, \qquad \vh_2=\om_2-\om_1 \,, \qquad \vh_3=-\om_2 \,.$$
\begin{figure}[ht]
  \centering
  \includegraphics[width=0.5\linewidth]{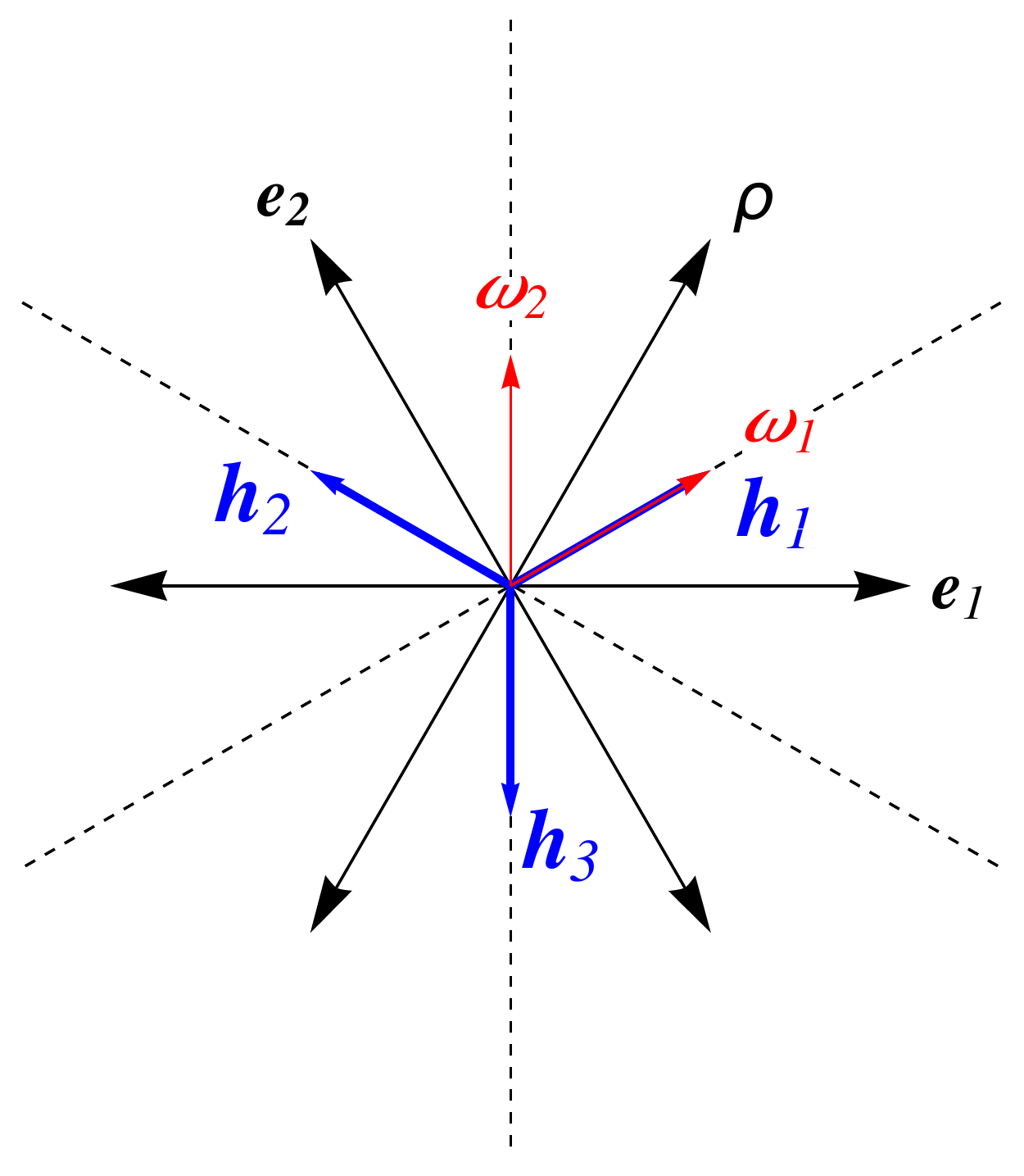}
  \caption{The generators of the root lattice $\mathcal{R}$ (in black) and the weight lattice $\mathcal{R}^*$ (in blue/red) for $\mathfrak{sl}_3$. The reflections of the Weyl group are the reflections with respect to the dashed lines.}
  \label{fig:roots-weights}
\end{figure}
\\
Let us review briefly the main lines of argument, which lead to the form of the CG action:
\begin{align}
  \mathcal{A}[\vect\phi] &= \int \frac{d^2x}{8\pi} \sqrt{|g|} \left[
    \partial_\mu\vect\phi \cdot \partial^\mu\vect\phi
    + 2i R(x) \vect Q \cdot \vect\phi + \sum_{\vect\alpha_{\rm scr}} \normal{e^{i\vect\alpha_{\rm scr} \cdot \vect\phi}}
    \right] \,, \label{eq:A} \\
  \vect\phi &\equiv \vect\phi + 2\pi b \mathcal{R} \,. \label{eq:phi=phi+R}
\end{align}
In the above expression, $\vect\phi$ is a two-component bosonic field, $g$ is the metric, $R(x)$ is the scalar curvature, $\vect Q$ is a constant vector called the ``background charge'', and $\{\vect\alpha_{\rm scr}\}$ is the set of possible screening charges, {\it i.e.} the charges compatible with the periodicity~\eqref{eq:phi=phi+R}, and with a marginal scaling dimension $h=\hb=1$. The parameter $b$ fixes the compactification scale. The central charge of the CFT is given by
\begin{equation} \label{eq:cc-CG}
  c = 2 - 12 \vect{Q}^2 \,,
\end{equation}
and the vertex operators and their conformal dimensions are:
\begin{equation} \label{eq:vertex}
  V_\alpha = \normal{e^{i\vect\alpha \cdot \vect\phi}} \quad \,,
  \qquad h_{\vect\alpha} = \frac{1}{2} \vect\alpha \cdot (\vect\alpha-2\vect Q) \,.
\end{equation}
The construction~\cite{Kondev-FPL} of the scaling theory (\ref{eq:A}--\ref{eq:phi=phi+R}) from the lattice FPL model goes as follows.
\begin{enumerate}
\item Each closed loop is assigned an alternation $(\vh_1,\vh_3,\vh_1 \dots)$ or $(\vh_3,\vh_1,\vh_3 \dots)$ starting from a particular edge, whereas the empty edges are assigned the label $\vh_2$. If the loop weight is written
  \begin{equation} \label{eq:n(lambda)}
    n = 2\cos 6\lambda \,,
  \end{equation}
  it can be distributed into local factors $e^{\pm i \lambda}$ for each loop turn of angle $\pi/3$, according to whether the loop turns right or left, and alternates from $\vh_1$ to $\vh_3$ or the reverse. Such a labelling of the edges thus leads to a vertex model on the hexagonal lattice, where every vertex is adjacent to exactly one of each of the labels $\vh_1,\vh_2,\vh_3$. On the dual triangular lattice, one then defines height variables, with local steps given by the edge labels: we take the convention that the step $2\pi b \vh_j$ is added to the neighbouring height when going clockwise around a Y-shaped vertex. Consider the height difference introduced by a closed loop in this construction: depending on the individual labelling $(\vh_1,\vh_3,\vh_1 \dots)$ or $(\vh_3,\vh_1,\vh_3 \dots)$ of the loop, it will be of the form $2\pi b (\vh_1+k\vh_2)$ or $2\pi b (\vh_3+k\vh_2)$, with integer $k$.

  After coarse-graining, if the surface is flat, the phase factors $e^{\pm i \lambda}$ give rise to the first term in the action \eqref{eq:A}.

\item The second term in \eqref{eq:A} ensures that the loops enclosing a region of non-zero curvature ({\it i.e.} with a total turning angle different from $\pm 2\pi$) get the proper Boltzmann weight $n$ : see also~\cite{FodaNienhuis89}. More precisely, if we consider a loop with turning angle $2\pi-\delta$, depending on its labelling the curvature term will insert a phase factor $\exp(-i\delta \, b\vh_1 \cdot \vect Q)$ or $\exp(-i\delta \, b\vh_3 \cdot \vect Q)$, which must correspond respectively to $\exp(- 6i\lambda \, \frac{\delta}{2\pi})$ and $\exp(+ 6i\lambda \, \frac{\delta}{2\pi})$. The correct choice is thus
  \begin{equation} \label{eq:Q(lambda)}
    \vect Q= q \vect\rho \,,
    \qquad \text{with} \quad q= \frac{6\lambda}{2\pi b} \,.
  \end{equation}

\item The vertex charges in the interaction terms of \eqref{eq:A} arise from the Fourier expansion of a $2\pi b \mathcal{R}$-periodic interaction potential in the discrete model:
\begin{equation} \label{eq:potential}
  \mathcal{V}[\vect \phi] = \sum_{\vect n \in \mathcal{R}^*} \kappa(\vect n) \, \exp(i \vect n \cdot \vect\phi/b) \,,
\end{equation}
where $\kappa(\vect n)$ is a constant amplitude.
Only the marginal terms in this expansion may appear in the action. The relevant terms act as ``energy-like'' operators, and the associated coupling constants in the lattice model must be fine-tuned so that the model is at its critical point. Finally, the irrelevant operators are suppressed in the scaling limit.

Let us first compute the conformal dimension of a generic field $V_{(n_1 \om_1 + n_2 \om_2)/b}$:
\begin{equation}
  h_{(n_1 \om_1 + n_2 \om_2)/b} = \frac{1}{4b^2} \left[ \frac{1}{3}(n_1-n_2)^2 + (n_1+n_2-2qb)^2 \right] - q^2 \,.
\end{equation}
In \cite{Kondev-FPL} it is argued that the terms appearing in \eqref{eq:potential} are actually of the form $V_{k\vect\rho/b}$ with $k\in \mathbb{Z}$, since they correspond to a ``chirality operator'' coupling to $(\vh_1-\vh_3)$ and \emph{not} to $\vh_2$. Let $k^*$ be the value for which $V_{k^*\vect\rho/b}$ is marginal. One then has
$$q = \frac{1}{2} \left( \frac{k^*}{b} - \frac{b}{k^*} \right) \,,$$
and the dimension of any $V_{k\vect\rho/b}$ reads:
$$h_{k\vect\rho/b} = 1 + (k/b^2+1/k^*)(k-k^*) \,.$$
In order for every $V_{k\vect\rho/b}$ to be irrelevant for all $k \notin \{0,k^*\}$, one must set $k^*=1$ and $0<b<1$.
This yields the relation: 
\begin{equation} \label{eq:Q(b)}
  \vect{Q} = \frac{1}{2} \left( \frac{1}{b} - b \right) \, \vect\rho \,,
  \qquad 0<b<1 \,,
\end{equation}
and the only screening charge in \eqref{eq:A} is then $V_{\vect\rho/b}$. Note that some relevant vertex operators are present in the spectrum of the transfer matrix, like for instance $V_{\pm \vh_2/b}$. They play the role of ``energy-like'' operators (see above). Moreover, in the limit $b \to 1$, the symmetry is enhanced to a Kac-Moody algebra. The six operators $V_{\pm \vect\rho}, V_{\pm \ve_1}, V_{\pm \ve_2}$ then become marginal.

\end{enumerate}

By combining \eqref{eq:n(lambda)}, \eqref{eq:Q(lambda)} and \eqref{eq:Q(b)}, one finds the relation between the loop fugacity $n$ and the compactification scale $b$ :
\begin{equation}
  n = -2 \cos \pi b^2 \,,
  \qquad 0<b<1 \,.
\end{equation}
It is important to notice that, despite their strong similarity, the scaling theory (\ref{eq:A}--\ref{eq:Q(b)}) for the FPL model and the imaginary $\mathfrak{sl}_3$ Toda CFT \cite{ImToda} are different. In particular, the screening charges in the imaginary Toda CFT are $V_{\ve_1/b}$ and $V_{\ve_2/b}$, and the background charge is $\vect{Q}_{\rm Toda} = (1/b-b)\vect\rho$. In \cite{DEI-FPL} we described a possible modification of the FPL on the cylinder, whose spectrum contains degenerate operators under the $W_3$ algebra, but in the present paper we will simply deal with the unadulterated FPL model as in, {\it e.g.} \cite{Kondev-FPL}.

If we write the components of $\vect\phi$ as $\vect\phi = \phi_1 \vect\rho + \phi_2 \vh_2$, the CG action becomes
\begin{align} \label{eq:A-decomp}
  \mathcal{A}[\vect \phi] = \int \frac{d^2x}{8\pi} \sqrt{|g|} \left[
    2\partial_\mu\phi_1 \partial^\mu\phi_1
    + 4i q R(x) \phi_1 + \normal{e^{2i\phi_1/b}}
    \right]
  + \frac{2}{3} \int \frac{d^2x}{8\pi} \sqrt{|g|} \partial_\mu\phi_2 \partial^\mu\phi_2 \,.
\end{align}
Hence, in the field configurations with no defects ({\it i.e.} the quantum fluctuations of $\vect\phi$ around the trivial classical solution of the equations of motion), the component $\phi_1$ is governed by an IL action with central charge
$$c_{\rm IL} = 1- 6(1/b-b)^2 \,, $$
and the $\phi_2$ component behaves as a free boson, and the two are decoupled. However, the full spectrum of the FPL model is determined by the set of all defect configurations allowed by the condition \eqref{eq:phi=phi+R}, which couples the components $\phi_1$ and $\phi_2$.

\subsection{Partition function and two-point functions}
\label{sec:2pt}

Through the procedure described in \secref{CG-action}, each individual closed loop, depending on its labelling, introduces a step of $2\pi b \vh_1$ or $2\pi b \vh_3$ for the field $\vect\phi$ between its inner and outer regions, and hence the loop fugacities may be adjusted by introducing proper factors of the form $e^{i \vect\alpha \cdot \vect\phi}$.

Let us first consider the partition function \eqref{eq:Z} on the Riemann sphere. The curvature term with $\sqrt{|g|}R(x) =8\pi \delta(x-x_\infty)$ in \eqref{eq:A} introduces a total U$(1)$ charge of $-2\vect{Q}$. Hence, in the absence of screening charges, any vertex correlation function $\aver{V_{\vect \alpha_1} \dots V_{\vect \alpha_N}}$ should satisfy the neutrality condition
\begin{equation} \label{eq:neutrality}
  \vect \alpha_1+ \dots +\vect \alpha_N = 2\vect Q \,.
\end{equation}
In particular, the partition function is represented as a one-point function in the CG formalism:
\begin{equation}
  Z_{\rm sphere} \propto \aver{V_{2\vect Q}(x_0)} \,,
\end{equation}
where $x_0$ is an arbitrary point. This can be interpreted as the two-point function
$$Z_{\rm sphere} \propto \aver{e^{2i\vect{Q} \cdot[\vect\phi(x_0)-\vect\phi(x_\infty)]}}_0$$
for the free-field action $\mathcal{A}_0[\vect\phi] = \int \frac{d^2x}{8\pi} \sqrt{|g|} \partial_\mu\vect\phi \cdot \partial^\mu\vect\phi$. Hence, the effect of the vertex operator $V_{2\vect Q}$ is to insert an additional factor $e^{-4i\pi b \vect{Q} \cdot \vh_1}$ or $e^{-4i\pi b \vect{Q} \cdot \vh_3}$ to each closed loop separating $x_0$ from $x_\infty$, depending on the labelling of the loop. When combining with the local $e^{\pm i\lambda}$ factors encoded in $\mathcal{A}_0$, the overall Boltzmann weight for such a loop is thus
$$e^{6i\lambda-4i\pi b \vect{Q} \cdot \vh_1} + e^{-6i\lambda-4i\pi b \vect{Q} \cdot \vh_3} = 2 \cos 6\lambda = n \,,$$
and thus all loops are assigned the same weight $n$, as required for~\eqref{eq:Z}.

When two points $x_1,x_2$ are marked on the sphere, there are two homotopy classes of closed loops: the contractible loops ({\it i.e.} those which do not separate the two marked points), and the non-contractible ones. We define the two-point correlation function by changing the fugacity of non-contractible loops to $n'$ in \eqref{eq:Z}:
\begin{equation} \label{eq:2pt}
  \mathcal{G}_{n'}(x_1,x_2) = \frac{1}{Z_{\rm FPL}} \sum_{\text{config. $C$}} n^{\ell_0(C,x_1,x_2)}
  \, (n')^{\ell'(C,x_1,x_2)} \,,
\end{equation}
where $\ell_0(C,x_1,x_2)$ [resp. $\ell'(C,x_1,x_2)$] is the number of contractible (resp. non-contractible) loops on the sphere punctured at $x_1$ and $x_2$. Through the same line of arguments as above, one can easily express this correlation function as a two-point function of vertex operators:
\begin{equation}
  \mathcal{G}_{n'}(x_1,x_2) \propto \aver{V_{\vect Q + p\vect\rho}(x_1) V_{\vect Q - p\vect\rho}(x_2)} \,,
  \qquad \text{with} \quad n'= 2 \cos 2\pi bp \,.
\end{equation}
The conformal dimension of $V_{\vect{Q} \pm p\vect\rho}$ is
\begin{equation} \label{eq:Delta_p}
  \Delta_p = h_{\vect{Q} \pm p\vect\rho} = p^2 - \frac{1}{4} \left(\frac{1}{b} - b\right)^2 \,.
\end{equation}
Note that a given value of $n'$ corresponds to infinitely many values of $p$, differing by multiples of $b^{-1}$. In the scaling limit, only the most relevant value remains: $|2bp| \leq 1$. We shall denote by $A_{n'}$ the non-universal amplitude of the two-point function, so that, in the scaling limit:
\begin{equation}
  \mathcal{G}_{n'}(x_1,x_2) = \frac{A_{n'}}{|x_1-x_2|^{4\Delta_p}} \,.
\end{equation}
The values $p= \pm q$ correspond to $n'=n$, and the associated vertex operators in the CG picture are $V_0$ and $V_{2\vQ}$, with conformal dimension $\Delta=0$.

\subsection{Three-point functions}
\label{sec:3pt}

\begin{figure}[ht]
  \centering
  \includegraphics[width=0.8\linewidth]{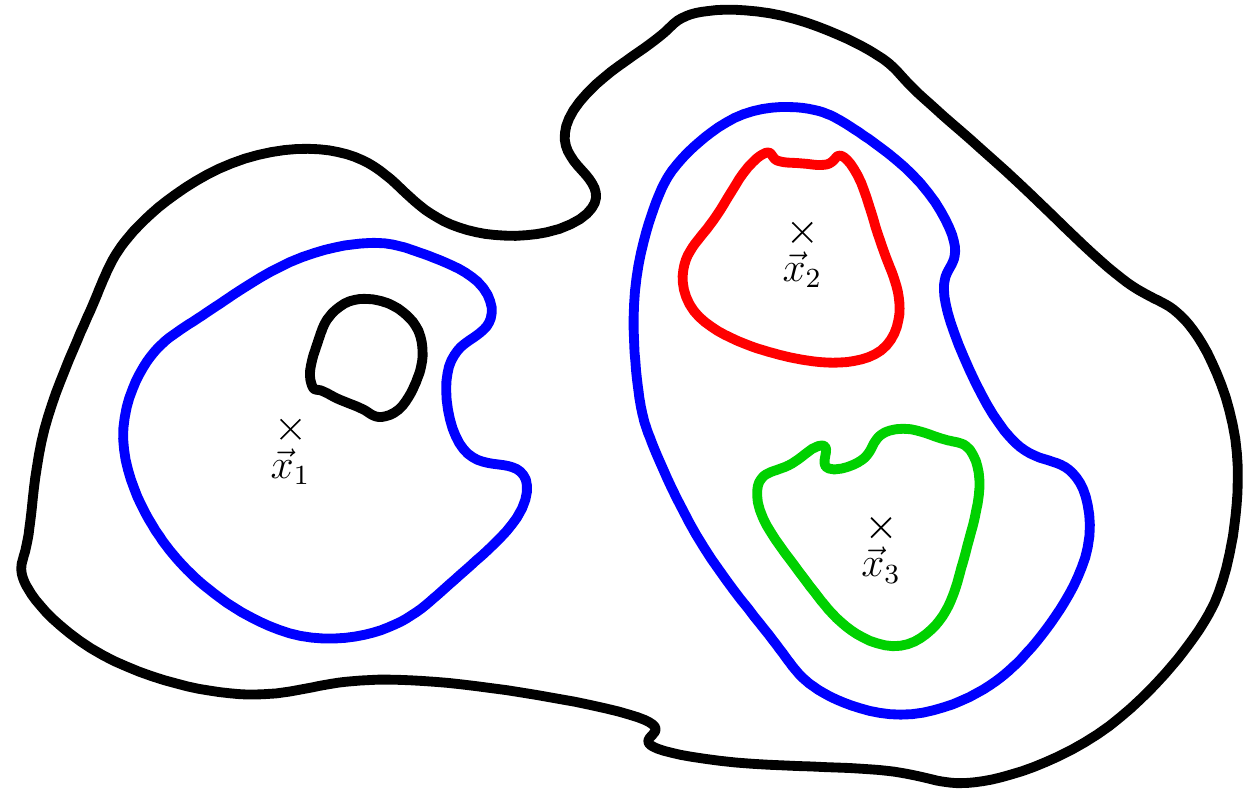}
  \caption{Classes of loops in the three-point function $\mathcal{G}_{n_1,n_2,n_3}$. The loops are coloured according to their homotopy on the Riemann sphere with three punctures at $x_1, x_2, x_3$. Contractible loops (in black) have fugacity $n$. Non-contractible loops, in blue, red, green, have fugacity $n_1,n_2,n_3$, respectively.}
  \label{fig:3pt}
\end{figure}

For the FPL model on the Riemann sphere, we define a family of three-point functions in analogy to the above two-point functions. Let $(x_1,x_2,x_3)$ be three given points on the sphere, and let three parameters $(n_1,n_2,n_3)$ be parameterised by
\begin{equation} \label{eq:nj-pj}
  n_j = 2 \cos 2\pi bp_j \,, \qquad \text{for} \quad j=1,2,3 \,.
\end{equation}
We consider the following object (see \figref{3pt}):
\begin{equation} \label{eq:3pt}
  \mathcal{G}_{n_1,n_2,n_3}(x_1,x_2,x_3) = \frac{1}{Z_{\rm FPL}} \sum_{\text{config. $C$}} n^{\ell_0(C,x_1,x_2, x_3)} \, \prod_{j=1}^3 n_j^{\ell_j(C,x_1,x_2, x_3)} \,,
\end{equation}
where $\ell_0(C,x_1,x_2,x_3)$ is the number of contractible loops on the sphere punctured at $x_1$, $x_2$ and $x_3$, and $\ell_j(C,x_1,x_2,x_3)$ is the number of loops which separate $x_j$ from the two other elements of $\{x_1,x_2,x_3\}$.

From general CFT arguments, if we admit that, in the scaling limit, $\mathcal{G}_{n_1,n_2,n_3}$ becomes the three-point function of primary operators, then it must take the form:
\begin{align} \label{eq:G=C}
  \mathcal{G}_{n_1,n_2,n_3}(x_1,x_2,x_3) = \frac{\sqrt{A_{n_1} A_{n_2} A_{n_3}} \ \mathcal{C}_n(n_1,n_2,n_3)}{|x_{12}|^{2(\Delta_1+\Delta_2-\Delta_3)}|x_{13}|^{2(\Delta_1+\Delta_3-\Delta_2)}|x_{23}|^{2(\Delta_2+\Delta_3-\Delta_1)}} \,,
\end{align}
where we have used the short-hand notations $x_{ij} = x_i-x_j$, and $\Delta_j=\Delta_{p_j}$.

In the particular case when $p_1,p_2,p_3$ satisfy the neutrality condition $p_1+p_2+p_3= -q$, through an argument analog to that of \secref{2pt}, one can indeed interpret $\mathcal{G}_{n_1,n_2,n_3}$ as the following three-point function of vertex operators:
\begin{equation} \label{eq:G=<VVV>}
  \mathcal{G}_{n_1,n_2,n_3}(x_1,x_2,x_3) \propto \aver{V_{\vQ+p_1 \vect\rho}(x_1)V_{\vQ+p_2 \vect\rho}(x_2)V_{\vQ+p_3 \vect\rho}(x_3)} \,.
\end{equation}
The main assumption of this paper is that this identification still holds when the neutrality condition is \emph{not} satisfied, so that for generic $(n_1,n_2,n_3)$, the constant $\mathcal{C}_n(n_1,n_2,n_3)$ is given by the three-point amplitude of vertex operators:
\begin{equation} \label{eq:C123}
  C_b(p_1,p_2,p_3) = \lim_{R \to \infty} \left[|R|^{4\Delta_1}\aver{V_{\vQ+p_1 \vect\rho}(R)V_{\vQ+p_2 \vect\rho}(1)V_{\vQ+p_3 \vect\rho}(0)} \right] \,.
\end{equation}

From the decomposition~\eqref{eq:A-decomp} of the CG action, the three-point amplitude~\eqref{eq:C123} is simply given by the IL CFT result~\cite{Schomerus03,Zamo05,KostovPetkova06}:
\begin{equation} \label{eq:C-IL}
  C_b(p_1,p_2,p_3) = \frac{\mathcal{A}_b \Upsilon_b \left(\mu_b+p_1+p_2+p_3 \right) \Upsilon_b \left(\mu_b+p_{12}^3 \right) \Upsilon_b \left(\mu_b+p_{23}^1 \right) \Upsilon_b \left(\mu_b+p_{13}^2 \right)} {\sqrt{\prod_{j=1}^3 \Upsilon_b(b+2p_j)\Upsilon_b(b-2p_j)}} \,,
\end{equation}
where $\mu_b=\frac{1}{2}(b+b^{-1})$, $p_{ij}^k = p_i+p_j-p_k$, and the function $\Upsilon_b$ is given by
\begin{equation}
  \Upsilon_b(x) = \exp \int_0^\infty \frac{dt}{t} \left[
    (\mu_b-x)^2 e^{-t} - \frac{\sinh^2(\mu_b-x)\frac{t}{2}}{\sinh \frac{bt}{2} \sinh \frac{t}{2b}}
    \right]
\end{equation}
for $0< \mathrm{Re}(x)<2 \mu_b$, and satisfies the relations
\begin{align}
  \Upsilon_b(x+b) = \frac{\Gamma(bx)}{\Gamma(1-bx)} \, b^{1-2bx} \, \Upsilon_b(x) \,,
  \qquad \text{and} \qquad
  \Upsilon_b(2\mu_b-x) = \Upsilon_b(x)
\end{align}
for any $x \in \mathbb{R}$. The normalisation factor in \eqref{eq:C-IL} is such that for any $p$, one has $C(p,p,q)=1$, corresponding to the three-point function $\aver{V_{\vQ+p \vect\rho}(\infty) V_{\vQ-p \vect\rho}(1) V_{0}(0)}$:
\begin{equation}
  \mathcal{A}_b = \frac{\Upsilon_b(2b-b^{-1})^{1/2}}{\Upsilon_b(b)^{3/2}} \,.
\end{equation}

One should note that, even though the conformal dimension associated to $p=\pm q$ is $\Delta=0$, the amplitude $C_b(p_1,p_2,\pm q)$ does not vanish for $p_1 \neq p_2$ :
\begin{equation}
  C_b(p_1,p_2,\pm q) = \frac{\Upsilon_b(b+p_1+p_2) \Upsilon_b(b-p_1-p_2) \Upsilon_b(b+p_1-p_2) \Upsilon_b(b-p_1+p_2)} {\Upsilon_b^2(b) \sqrt{\prod_{j=1}^2 \Upsilon_b(b+2p_j)\Upsilon_b(b-2p_j)}} \,,
\end{equation}
This corresponds to the fact that the function $\mathcal{G}_{n_1,n_2,n}$ is genuinely a three-point function. For instance, a loop enclosing only $x_1$ is assigned the weight $n_1$, whereas a loop enclosing both $x_1$ and $x_3$ will get the weight $n_2$. Hence, the vertex operator $V_0$ acts as a ``marking operator'' in the loop model, with a non-trivial effect despite its vanishing dimension.

\section{Numerical analysis}
\label{sec:num}

\subsection{Correlation functions on the cylinder}
\label{sec:cyl}

Although this is based on standard CFT arguments, for completeness, we expose how the correlation functions on the cylinder are related to those on the Riemann sphere. Consider an cylinder of infinite length, and finite circumference $L$, which we represent by the region $\{ w \in \mathbb{C} \,, 0 \leq \mathrm{Im}(w) \leq L \}$ with periodic boundary conditions in the imaginary direction. This surface is mapped to the Riemann sphere by the function $w \mapsto z = \exp(2\pi w/L)$. Using conformal covariance, one has:
\begin{align}
  & \aver{\Phi_1(w_1,\wb_1) \Phi_2(w_2,\wb_2) \dots  \Phi_N(w_N,\wb_N)}_{\rm cyl} \nn \\
  & \qquad = \prod_{j=1}^N \left(\frac{2\pi z_j}{L} \right)^{h_j}\left(\frac{2\pi \zb_j}{L} \right)^{\hb_j} \aver{\Phi_1(z_1,\zb_1) \Phi_2(z_2,\zb_2) \dots  \Phi_N(z_N,\zb_N)}_{\rm sph} \,,
\end{align}
where $z_j=\exp(2\pi w_j/L)$, and $\Phi_j$ is a primary operator of conformal dimensions $(h_j,\hb_j)$. In particular, by sending $w_1$ and $w_N$ to the ends of the cylinder, we obtain:
\begin{align*}
  & \frac{\aver{\Phi_1(M,\bar M) \Phi_2(w_2,\wb_2) \dots  \Phi_{N-1}(w_{N-1},\wb_{N-1}) \Phi_N(-M,-\bar M)}_{\rm cyl}}
  {\sqrt{\aver{\Phi_1(M,\bar M)\Phi_1(-M,-\bar M)}_{\rm cyl} \aver{\Phi_N(M,\bar M)\Phi_N(-M,-\bar M)}_{\rm cyl}}} \\
  & \qquad \mathop{\longrightarrow}_{M \to +\infty} \prod_{j=2}^{N-1} \left(\frac{2\pi z_j}{L} \right)^{h_j}\left(\frac{2\pi \zb_j}{L} \right)^{\hb_j} \aver{\Phi_1(\infty) \Phi_2(z_2,\zb_2) \dots \Phi_{N-1}(z_{N-1},\bar z_{N-1}) \Phi_N(0)}_{\rm sph} \,.
\end{align*}

The basic ingredients of the transfer matrix formalism are the Euclidian time evolution operator (or transfer matrix) $t_L$ on the cylinder, and a bilinear form (usually called ``scalar product'' even when it is not positive definite) $\aver{\cdot,\cdot}$, such that the transfer matrix is self-adjoint: $\aver{u,t_Lv} = \aver{t_Lu,v}$ for any vectors $u,v$. The operator-state correspondence assumes that a primary operator $\Phi_j$, when acting on the conformally invariant state $\psi_{\rm vac}$ (the ``vacuum'' state), produces an eigenstate $\psi_j$ :
$$\Phi_j(0) \cdot \psi_{\rm vac} \propto \psi_j \,.$$
The cylinder correlation functions are represented as overlaps of eigenstates $\psi_j$, with some insertions of operators $\Phi_k$. From the above, we get
\begin{align}
  & \frac{\aver{\psi_1, \Phi_2(w_2,\bar w_2) \dots  \Phi_{N-1}(w_{N-1},\bar w_{N-1}) \psi_N}}
  {\sqrt{\aver{\psi_1,\psi_1}\aver{\psi_N,\psi_N}}}
  \times \prod_{j=2}^{N-1} \frac{\sqrt{\aver{\psi_j,\psi_j} \aver{\psi_{\rm vac},\psi_{\rm vac}}}}
       {\aver{\psi_j,\Phi_j(0) \psi_{\rm vac}}} \nn \\
  & \qquad\qquad\qquad = \aver{\Phi_1(\infty) \Phi_2(z_2,\zb_2) \dots \Phi_{N-1}(z_{N-1},\bar z_{N-1}) \Phi_N(0)}_{\rm sph} \,. \label{eq:overlap}
\end{align}
The second factor in~\eqref{eq:overlap} is particularly important for lattice computations, where the normalisation of the operators $\Phi_j$ depends on the microscopic details of the lattice model. The above arguments also predict the scaling:
\begin{equation} \label{eq:scaling-Phi_j}
  \frac{\aver{\psi_j,\Phi_j(0) \psi_{\rm vac}}}
       {\sqrt{\aver{\psi_j,\psi_j} \aver{\psi_{\rm vac},\psi_{\rm vac}}}}
  \sim \mathrm{const} \times L^{-h_j-\hb_j} \,.
\end{equation}

Correlation functions of primary operators in a given CFT may be expressed in the critical lattice model, as long as, for each primary field, one is able to identify the eigenstate and lattice operator which scale to the corresponding primary state and operator. We shall then use \eqref{eq:overlap} to check the identification of lattice correlation functions with CFT $N$-point functions. In particular, in the case of two-point functions of scalar primary operators, we have the relation:
\begin{equation}
  \frac{\aver{\psi_j,\psi_j} \, \aver{\psi_{\rm vac},\Phi_j(0) \Phi_j(w, \wb) \psi_{\rm vac}}}
       {\aver{\psi_j,\Phi_j(0)\psi_{\rm vac}}^2}
       = \frac{1}{\left| 2 \sinh \frac{\pi w}{L} \right|^{4h_j}} \,.
\end{equation}
For the three-point function of scalar primary operators, one may use
\begin{equation} \label{eq:3pt-cyl1}
  \frac{\aver{\psi_1,\Phi_2(0) \psi_3}}{\aver{\psi_2,\Phi_2(0)\psi_{\rm vac}}}
  \sqrt{\frac{\aver{\psi_2,\psi_2}\aver{\psi_{\rm vac},\psi_{\rm vac}}}{\aver{\psi_1,\psi_1}\aver{\psi_3,\psi_3}}}
  = C_{123} \,,
\end{equation}
or
\begin{equation} \label{eq:3pt-cyl2}
  \frac{\aver{\psi_1,\Phi_2(0) \Phi_3(w, \wb) \psi_{\rm vac}}}
       {\aver{\psi_2,\Phi_2(0)\psi_{\rm vac}}\aver{\psi_3,\Phi_3(0)\psi_{\rm vac}}}
       \sqrt{\frac{\aver{\psi_2,\psi_2}\aver{\psi_3,\psi_3}\aver{\psi_{\rm vac},\psi_{\rm vac}}}{\aver{\psi_1,\psi_1}}}
       = \frac{C_{123}}{\left| 2 \sinh \frac{\pi w}{L} \right|^{2(h_2+h_3)}} \,.
\end{equation}

\subsection{Transfer matrix and scalar product for the FPL model}
\label{sec:tm}

\begin{figure}[ht]
  \centering
  \begin{tabular}{ccc}
    \includegraphics{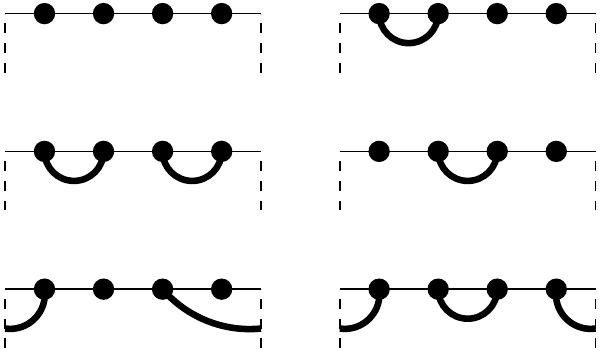} && \includegraphics{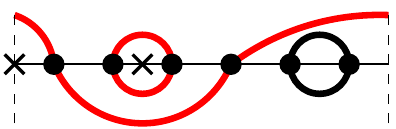} \\
    \\
    (a) & \qquad\qquad & (b)
  \end{tabular}
  \caption{Link patterns and scalar product for the FPL model. (a) Some link pattern states for the transfer matrix with $L=4$. (b) The modified scalar product $\aver{\cdot, \Phi_{n'}(0)\Phi_{n'}(\ell) \cdot}$ with $\ell=2$ and $L=6$. The black (resp. red) loops get a weight $n$ (resp. $n'$).}
  \label{fig:states}
\end{figure}

The transfer matrix formalism for loop models is well described, {\it e.g.} in~\cite{BloteNienhuis89}. For convenience, we rotate the system by 90$^\circ$, so that the transfer matrix acts in the vertical direction. The basis states are labelled by link patterns connecting $L$ points (see \figref{states}, and the transfer matrix $t_L$ encodes the Boltzmann weights for one row with periodic boundary conditions. The scalar product of two link patterns $a$ and $b$ is given by $\aver{u,v} = n^{m(u,v)}$, where $m(u,v)$ is the number of closed loops appearing when $u$ is connected to $\bar v$, the reflection of $v$ about the horizontal axis.

We shall denote by $\psi_{n'}$ the dominant eigenstate of the transfer matrix $t_L(n')$ with \emph{modified} periodic boundary conditions: the closed loops which wrap around the cylinder are assigned the weight $n'$ instead of $n$. In particular, the vacuum state is $\psi_{\rm vac} = \psi_n$. Similarly, we denote by $\Phi_{n'}$ the operator which sets to $n'$ the fugacity of loops surrounding it.

For instance, the ``equal-time'' two-point function $\aver{\psi_{\rm vac}, \Phi_{n'}(0) \Phi_{n'}(\ell) \psi_{\rm vac}}$ with $0<\ell<L$ is obtained by modifying the scalar product so that loops which separate the points $0$ and $\ell$ get the weight $n'$ instead of $n$. For any link patterns $u,v$, we set
\begin{equation}
  \aver{u,\Phi_{n'}(0) \Phi_{n'}(\ell) \, v} = n^{m_0(u,v,\ell)} \, (n')^{m'(u,v,\ell)} \,,
\end{equation}
where $m_0(u,v,\ell)$ [resp. $m'(u,v,\ell)$] is the number of loops which do not separate (resp. which separate) the points $0$ and $\ell$ when $u$ is connected to $\bar v$. When computing three-point functions such as $\aver{\psi_{n_1},\Phi_{n_2}(0) \psi_{n_3}}$ or $\aver{\psi_{n_1},\Phi_{n_2}(0)\Phi_{n_3}(\ell) \psi_{\rm vac}}$, it is necessary to modify the scalar product in order to assign the proper weight to each closed loop, according to \eqref{eq:3pt}.

\subsection{Numerical implementation and results}
\label{sec:plots}

In this section, we compare the theoretical prediction~\eqref{eq:C-IL} for the three-point amplitude in~\eqref{eq:G=C} with finite-size computations of three-point functions.
The eigenstates of the model are obtained by extracting the eigenvectors associated with the smallest real eigenvalues of the Hamiltonian operator, through a Krylov-Schur algorithm for sizes $L$ between $3$ and $15$. We use the range of loop fugacities $-2<n<2$, where the model is critical, and can be described by the CG action of \secref{CG}.
Let us give here some important implementation remarks:
\begin{itemize}
\item We only compute eigenvalues and eigenvectors for systems of size multiple of three. Indeed, if the system size $L$ is not a multiple of three, then from the local constraints of the FPL model, the configuration of the field $\vect\phi$ has a nonzero defect $\vect\phi \to \vect\phi + 2\pi b \vect m$ with $\vect{m} \in \mathcal{R}$ when going around the circumference. This corresponds, \textit{e.g.}, to an open path propagating along the cylinder.
\item The predictions (\ref{eq:3pt-cyl1}--\ref{eq:3pt-cyl2}) include some normalisation factors of the form $\sqrt{\aver{\psi,\psi}}$, but the ``scalar product'' defined in \secref{tm} is \emph{not} positive definite. Hence, even though the transfer matrix and the matrix elements of the operators $\Phi_j$ are real, one may obtain pure imaginary three-point amplitudes.
\item The agreement of our transfer-matrix calculations with the imaginary Liouville prediction~\eqref{eq:C-IL} through the relation~\eqref{eq:3pt-cyl1} is generally very good. The latter relation corresponds to placing two marked points of the correlation function $\mathcal{G}_{n_1,n_2,n_3}$~\eqref{eq:3pt}, say $x_1,x_3$, on the boundaries of the infinite cylinder, and the third point $x_2$ in the bulk of the cylinder. The relation~\eqref{eq:3pt-cyl2}, in which two marked points sit in the bulk and one is on the boundary, allows us to test both the value of the three-point amplitude \emph{and} the spatial dependence of $\mathcal{G}_{n_1,n_2,n_3}$ on the cylinder. In this case, the agreement is quite good, but suffers from large finite-size effects.
\end{itemize}

\paragraph{First example : $\mathcal{G}_{n',n',n'}$ as a function of $n'$.} We consider the FPL model at the dimer point, which corresponds to the fugacity $n=1$, and uniformly vary the weights of any loop surrounding one (and only one) of the three marked points. From~\eqref{eq:3pt-cyl1}, this may be computed as the ratio of overlaps:
\begin{equation}
  \mathcal{C}_n(n',n',n') = \frac{\aver{\psi_{n'},\Phi_{n'}(0) \psi_{n'}}}{\aver{\psi_{n'},\Phi_{n'}(0)\psi_{\rm vac}}}
  \sqrt{\frac{\aver{\psi_{\rm vac},\psi_{\rm vac}}}{\aver{\psi_{n'},\psi_{n'}}}}
   \,,
\end{equation}  
where $\psi_{n'}$ and $\Phi_{n'}$ are respectively the state and the operator corresponding to a loop fugacity $n'$, as defined in \secref{tm}. The results are shown in \figref{C111}, for sizes $L=3, \dots 12$.

The conjectured expression~\eqref{eq:C-IL} remains valid even for $n' \geq 2$, even though the corresponding ``momentum'' $p$ defined by $n'=2\cos 2\pi bp$ becomes pure imaginary, and is outside the spectrum of the imaginary Liouville CFT.

\begin{center}
  \includegraphics[width=\linewidth]{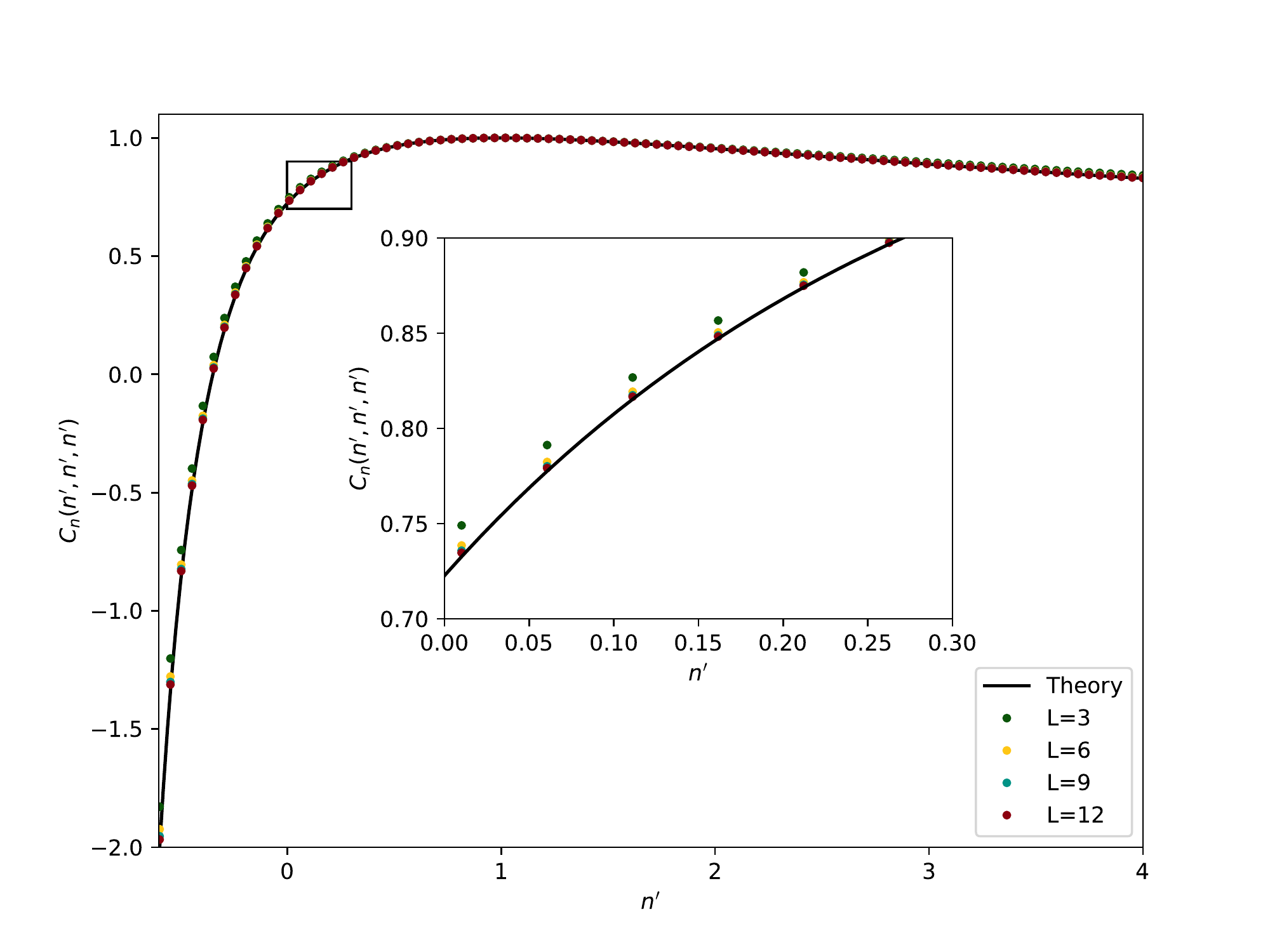}
  \captionof{figure}{The three-point amplitude $\mathcal{C}_n(n',n',n')$ as a function of $n'$. The black line corresponds to the theoretical value given by the IL formula \eqref{eq:C-IL}. The dots are numerical estimates on cylinders of circumference of $L$ sites. The insert shows a zoomed-in version which makes it possible to observe the convergence towards the theoretical value as $L$ increases.}
  \label{fig:C111}
\end{center}

\paragraph{Second example : three-point ``dual connectivity'' as a function of $n$.} 
Let us consider the function $\mathcal{G}_{0,0,0}$, where we keep the value of the modified loop fugacities fixed to $n_1=n_2=n_3=0$ (corresponding to $p=b^{-1}/4$), and we vary the value of $n$. This function has a nice statistical interpretation. To each FPL configuration we associate a subgraph of the dual triangular lattice consisting of the edges which do not cross a loop segment. Then, for $n \geq 0$, the function $\mathcal{G}_{0,0,0}(x_1,x_2,x_3)$ gives the probability that the three points $x_1,x_2,x_3$ are on the same connected component.

On this example, we see that the three-point amplitude can become pure imaginary, as predicted by the imaginary Liouville formula: indeed, at $n=-1$, we have $b=1/\sqrt{3}$ and $\mathcal{A}_b=0$ in \eqref{eq:C-IL}. At this point, the squared three-point amplitude $C_b^2(b^{-1}/4,b^{-1}/4,b^{-1}/4)$ changes sign. Correspondingly, we find that the norm of $\psi_{\rm vac}$ vanishes for $n=-1$. The results are shown in \figref{C000}.

The limit $n \to 0$ is also interesting, since it corresponds to Hamiltonian walks on the hexagonal lattice. In this case, the three-point amplitude is easily obtained as follows. First, since the scaling dimension $\Delta_{p=b^{-1}/4}$ vanishes, the two- and three-point functions should become position-independent in the scaling limit. The Hamiltonian walk separates the dual lattice into two regions of equal area, and hence one has $\mathcal{G}_{n'=0} = 1/2$ and $\mathcal{G}_{0,0,0} = 1/4$ in the scaling limit. Using the normalisation procedure described in \secref{3pt}, we get the three-point amplitude $\mathcal{C}_{n\to 0}(0,0,0)= 1/\sqrt{2}$. The corresponding computation through the imaginary Liouville formula, {\it i.e.} taking the limit $b \to 1/\sqrt{2}$ in the expression $C_b(b^{-1}/4,b^{-1}/4,b^{-1}/4)$, deserves a careful treatment, because at this point, both the numerator and the denominator of \eqref{eq:C-IL} vanish. The property $\Upsilon_b(x) \underset{x \to 0}{\sim} \Upsilon_b(b) \, x$ can be used to resolve this, indeed leading to $C_b(b^{-1}/4,b^{-1}/4,b^{-1}/4) \to 1/\sqrt{2}$.


\begin{center}
  \includegraphics[width=\linewidth]{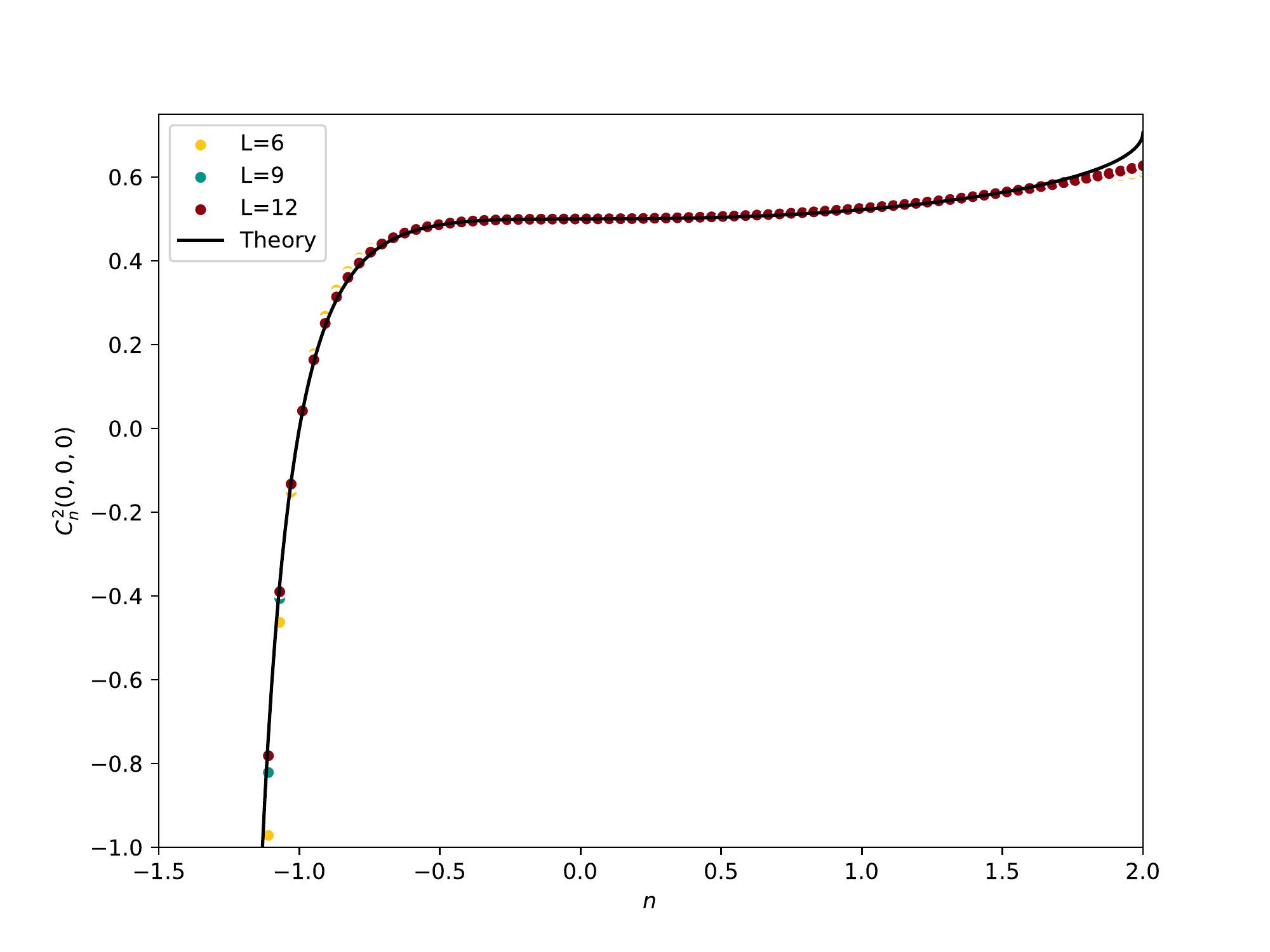}
  \captionof{figure}{The squared three-point amplitude $\mathcal{C}_n^2(0,0,0)$, related to the three-point dual connectivity, as a function of $n$. The black line corresponds to the theoretical value given by the IL formula \ref{eq:C-IL}. The dots are numerical estimates for different lattice sizes.
  } \label{fig:C000}
\end{center}

\paragraph{Marking operator.}
As already emphasized in \secref{3pt}, the amplitude $\mathcal{C}_n(n_1,n_2,n_3)$ remains non-trivial when one or several $n_j$'s are equal to $n$. The corresponding primary operator has dimension zero, but nevertheless has a non-trivial effect in correlation functions: we call it the marking operator, like in \cite{IJS16}. For instance, we have computed the three-point amplitude $\mathcal{C}_n(n',n,n)$ as a function of $n'$, in the dimer case $n=1$. The results are shown in \figref{marking}, and show excellent agreement with \eqref{eq:C-IL}.

\begin{center}
  \includegraphics[width=\linewidth]{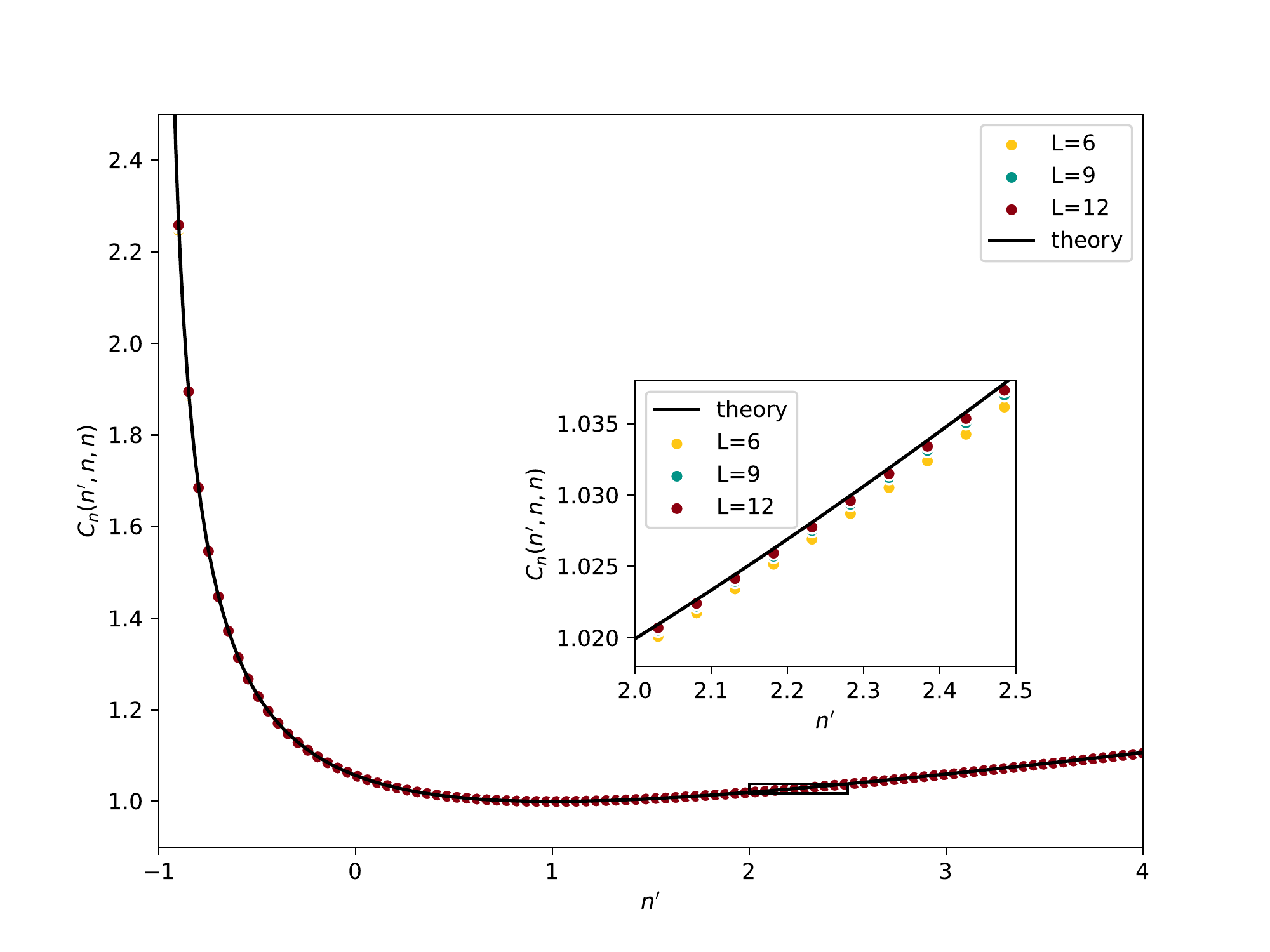}
  \captionof{figure}{The three-point amplitude $\mathcal{C}_n(n',n,n)$ as a function of $n'$ with $n=1$. The black line corresponds to the theoretical value given by the IL formula \ref{eq:C-IL}. The dots are numerical estimates for different lattice sizes. The insert shows a zoomed-in version which makes it possible to observe the convergence towards the theoretical value as $L$ increases. \label{fig:marking}}
\end{center}

\paragraph{Subleading primary states.}
The parameterisation~\eqref{eq:nj-pj} of the modified loop fugacities $n_j$ only determine the momenta $p_j$ up to the addition of multiples of $b^{-1}$. So far, we have focussed on the interval $|p_j|<b^{-1}/2$, corresponding to the most relevant state or operator associated to $n_j$. It is also possible to study higher values of $p_j$, which determine the subleading behaviour of $\mathcal{G}_{n_1,n_2,n_3}$ in the scaling limit. Setting $n=1$, in \figref{subleading}, we show the numerical computation for the three-point amplitude
\begin{equation} \label{eq:C-exc}
  \widetilde{\mathcal{C}}_n(n',n,n) =
  \frac{\smallaver{\widetilde \psi_{n'},\Phi_n(0) \psi_{\rm vac}}}{\aver{\psi_n,\Phi_n(0)\psi_{\rm vac}}}
  \sqrt{\frac{\aver{\psi_n,\psi_n}}{\smallaver{\widetilde \psi_{n'},\widetilde \psi_{n'}}}}
  = \frac{\smallaver{\widetilde \psi_{n'},\Phi_n(0) \psi_{\rm vac}}}
  {\sqrt{\smallaver{\widetilde \psi_{n'},\widetilde \psi_{n'}}\aver{\psi_{\rm vac},\psi_{\rm vac}}}} \,,
\end{equation}
where $\widetilde \psi_{n'}$ is the first excited state of the transfer matrix with non-contractible loop fugacity $n'$, and corresponds to a momentum $\widetilde p \in [b^{-1}/2, b^{-1}]$ through the relation $n'=2\cos 2\pi b \widetilde p$. In \eqref{eq:C-exc} we have used the fact that $\psi_{\rm vac}=\psi_n$ and $\aver{\psi_n,\Phi_n(0)\psi_{\rm vac}} = \aver{\psi_{\rm vac},\psi_{\rm vac}}$ to simplify the expression.

\begin{center}
  \includegraphics[width=\linewidth]{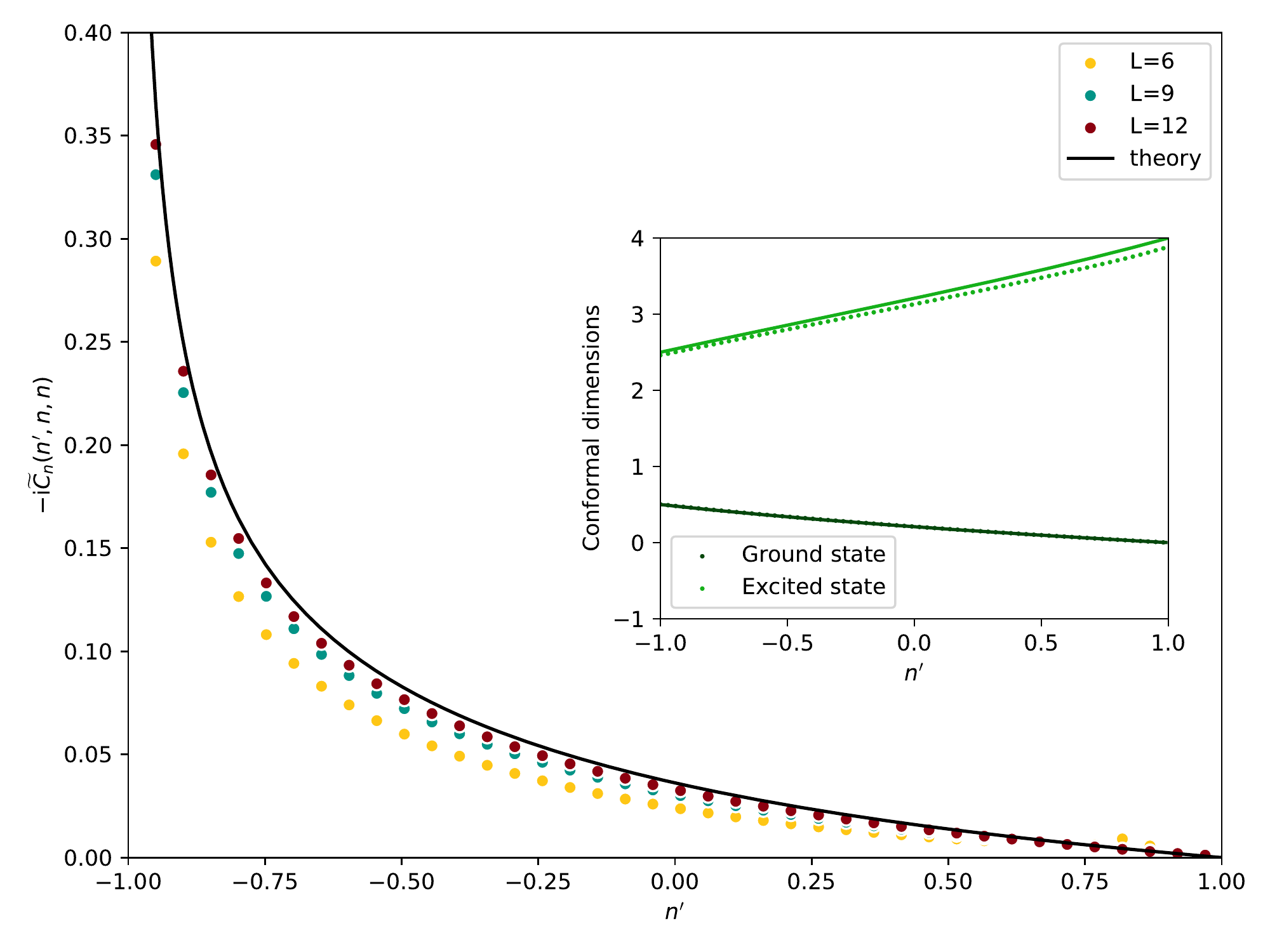}
  \captionof{figure}{The three-point amplitude $\widetilde{\mathcal{C}}_n(n',n,n)$ (defined in \ref{eq:C-exc}) as a function of $n'$, with fixed $n=1$. The black line corresponds to the theoretical value given by the IL formula \ref{eq:C-IL}. The dots are numerical estimates for different lattice sizes. The insert shows the conformal dimension as a function of $n'$ for the ground state $\psi_{n'}$ and for the excited state $\widetilde\psi_{n'}$. The full line corresponds to the theoretical formula \ref{eq:Delta_p} and the dotted line to their numerical counterpart ($L=12$). Details on how to extract the conformal weights from the eigenvalue of the Hamiltonian of the FPL model can be found in \cite{DEI-FPL}. \label{fig:subleading}}
\end{center}

\paragraph{Spatial dependence.}
We have tested the spatial dependence of the three-point function $\mathcal{G}_{n_1,n_2,n_3}$ on the cylinder, when two marked points are located on the same circumference, and the third point is at infinity. Our formula \eqref{eq:3pt-cyl2} gives the CFT prediction for this spatial dependence, with an overall factor given by \eqref{eq:C-IL}. We have considered the case $\mathcal{G}_{n',n,n}(0,i\ell,\infty)$, with $n=1$ and $n'=0$. The results are shown in \figref{2pt-bulk}. The agreement with CFT predictions is quite good, although it is altered by important finite-size effects.


\begin{center}
  \includegraphics[width=\linewidth]{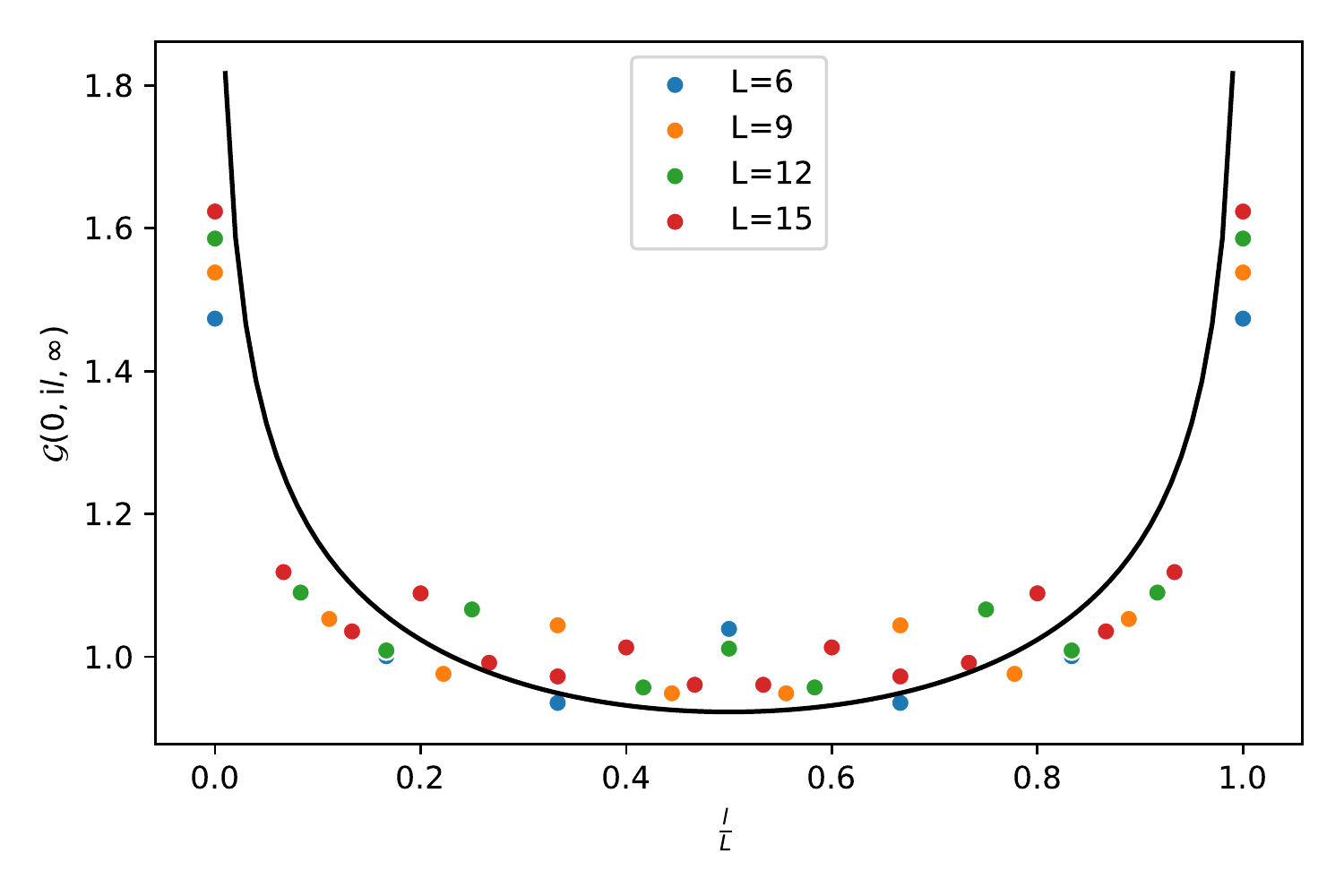}
  \captionof{figure}{The three-point function $\mathcal{G}_{n',n,n}(0,i\ell,\infty)$ on a cylinder of circumference $L$, as a function of $\ell/L$, for $n=1$ and $n'=0$.}
  \label{fig:2pt-bulk}
\end{center}




\paragraph{Comparison with other loop models.}

The correlation functions $\mathcal{G}_{n_1,n_2,n_3}$ and their three-point amplitude $\mathcal{C}_n(n_1,n_2,n_3)$ may be defined in any non-intersecting loop model. In \cite{IJS16}, it was shown that for the dilute O($n$) model and the Temperley-Lieb loop model, $\mathcal{C}_n(n_1,n_2,n_3)$ is given by the IL formula \eqref{eq:C-IL}. In \figref{diff_models}, we compare the three-point amplitude for generic values of the $n_j$'s in various loop models.

\begin{center}
  \includegraphics[width=\linewidth]{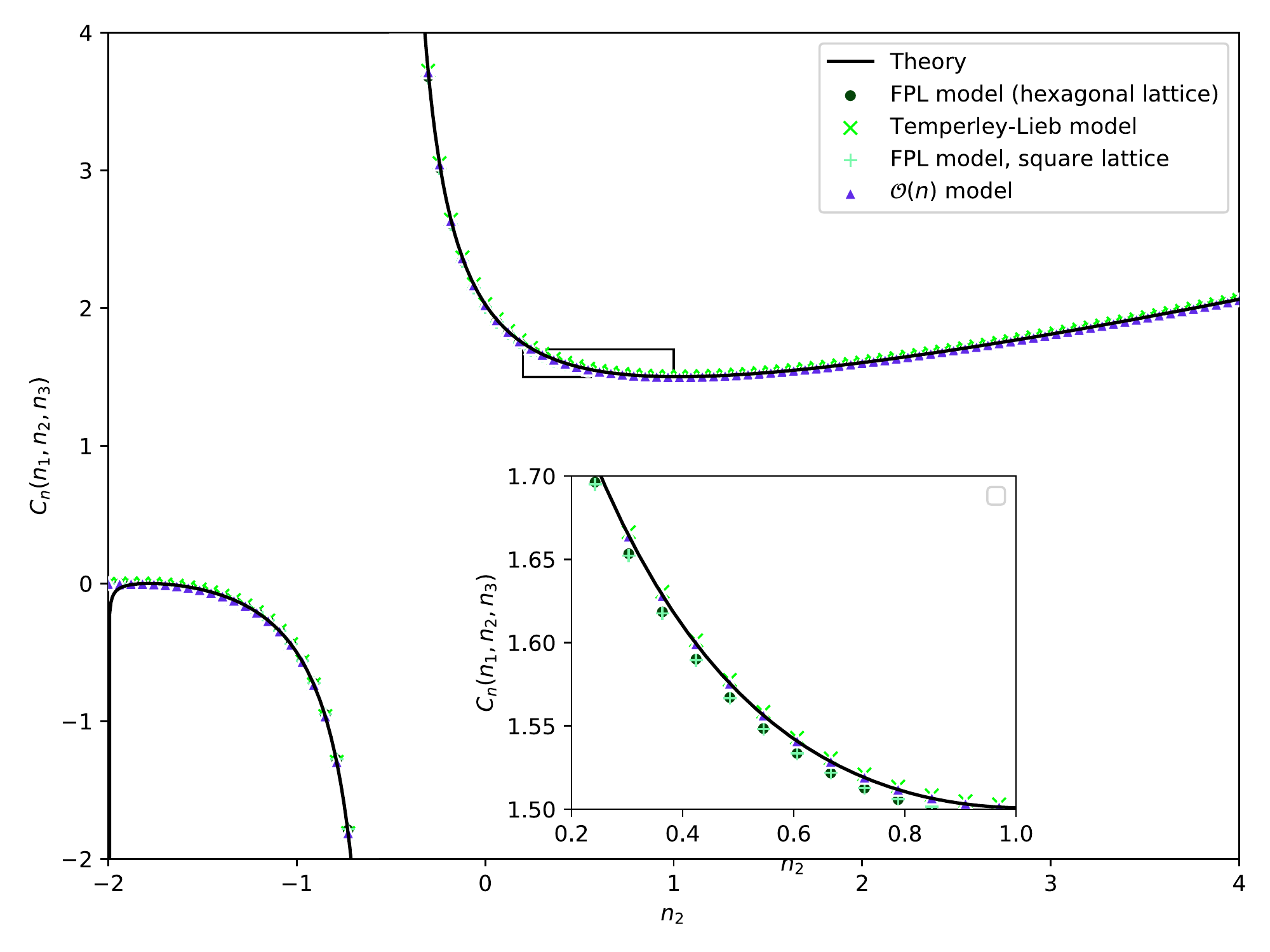}
  \captionof{figure}{The three-point amplitude $\mathcal{C}_n(n_1,n_2,n_3)$ as a function of $n_2$, with $n=0.5$, $n_1=0$, and $n_3=1.5$. The black line corresponds to the theoretical value given by the IL formula \ref{eq:C-IL}. The dots are numerical estimates for different models, on the same cylinder of diameter $L=12$. The definitions for the Temperley-Lieb, dilute $\mathcal{O}(n)$ and the FPL model on the square lattice can be found in \cite{batchelor1996critical, 0305-4470-22-9-028}.}
  \label{fig:diff_models}
\end{center}

\section{Conclusion}

It is known from the Coulomb gas approach\cite{Kondev-FPL} that the universal behavior of the Fully Packed Loop model on the honeycomb lattice is described by a two-component compact boson coupled to curvature, and with interaction terms in the form of screening charges. Up to compactification conditions, this theory can be interpreted as the tensor product of the imaginary Liouville theory and a free boson. In this paper we have studied a family of geometric two-point and three-point functions that probe the imaginary Liouville sector. Using a transfer matrix approach, we have checked numerically that the spatial dependence of these correlation functions is consistent with a correlation function of primary fields. The conformal dimension and structure constants obtained numerically are in very good agreement with those expected from the imaginary Liouville theory. 

More generally these geometric observables can be defined in any non-intersecting loop model. We conjecture that they are always described by the imaginary Liouville theory. We numerically substantiate this conjecture for various loop models.

\bibliographystyle{unsrt}
\bibliography{biblio}

\end{document}